\documentclass{pic2012}
\usepackage{subfigure}
\usepackage{amssymb,amsmath}
\def\runauthor{Tony M. Liss}
\def\shorttitle{Top Quark Properties}
\begin{document}
\title{Top Quark Properties
}

\author{Tony M. Liss}

\address{Department of Physics\\
University of Illinois\\
Urbana, Illinois 61801\\
 E-mail: tml@illinois.edu\\ {\rm On behalf of the CDF, D0, ATLAS and CMS Collaborations} }

\maketitle

\abstracts{I review the latest results on properties of the top
quark from the Tevatron and the LHC, including results measured in
$t\bar{t}$ and single-top events on the mass, width, couplings, and
spin correlations. }

\section{Introduction} With results from almost 9 fb$^{-1}$ of
$\bar{p} p$ collisions at $\sqrt{s}=1.96$ TeV at the Tevatron, and
up to 5 fb$^{-1}$ of $pp$ collisions at $\sqrt{s}=7$ TeV at the LHC,
measurements of top-quark properties are becoming quite precise.  I
will review the latest (as of early September, 2012) results from D0
and CDF at the Tevatron and CMS~\cite{CMSPaper} and
ATLAS~\cite{ATLASPaper} at the LHC. These results include the top
quark mass and width, the $t$ and $\bar{t}$ mass difference, top
quark couplings, and $t-\bar{t}$ spin correlations. The top quark
charge asymmetry is covered elsewhere~\cite{AsymTalk}. The event
selection and background evaluation are also covered
elsewhere~\cite{XsecTalk}.

\section{Top Quark Mass Measurements}
\begin{figure}[!thb]
 \centerline{\epsfxsize=2.9in\epsfbox{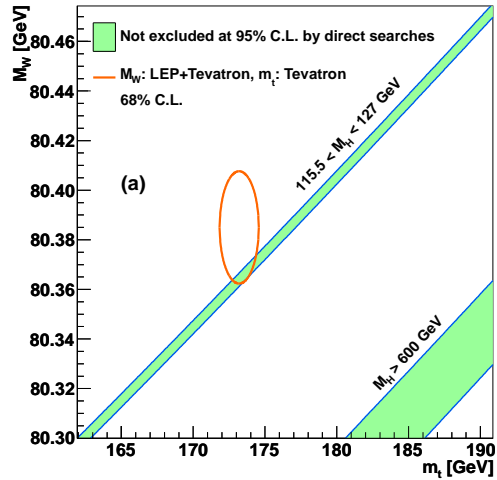}}
\caption[*]{ Top mass vs. $W$ boson mass showing bands of Standard
Model Higgs boson masses.  The ellipse shows the current Tevatron
averages.  The figure is from Reference~\cite{ref:MtMw}.}
\label{fig:MtMw}
\end{figure}

The top quark's mass, $m_t$, is its most fundamental property.  It
is also the most precisely measured property. Besides being a free
parameter in the Standard Model, the importance of the top quark
mass measurements stems from its role in electroweak radiative
corrections which yield a quadratic dependence of the $W$ boson
mass, $m_W$, on $m_t$, while $m_W$ only depends logarithmically on
$m_H$, the mass of the Higgs boson.  Precision measurements of $m_W$
and $m_t$ therefore probe the Standard Model Higgs boson mass, as
shown in Figure~\ref{fig:MtMw}, and with a precise measurement of
the Higgs boson mass probe the consistency of the Standard Model
itself.

The measurement of $m_t$ involves a number of challenges. The top
quark decays almost 100\% of the time to a $W$ boson and a $b$
quark. The $t\bar{t}$ final state, $W^+bW^-\bar{b}$ with each $W$
boson decaying to $\ell\nu$ or $q\bar{q'}$, involves six partons,
and these must be identified in the selected events and properly
assigned to the two $W$ bosons and the $b$ and $\bar{b}$ quarks.  In
addition to the parton assignment, the best possible energy and
momentum resolution must be achieved for each, and here the jet
energy scale (JES) is of paramount importance.

A large number of techniques have been used over the years for
measuring $m_t$.  Here I will discuss only those used in a few of
the most recent results.

\subsection{CDF Measurement in lepton+jets}
\label{sec:CDFLJ} 
\begin{figure}[!thb]
 \centerline{\epsfxsize=2.9in\epsfbox{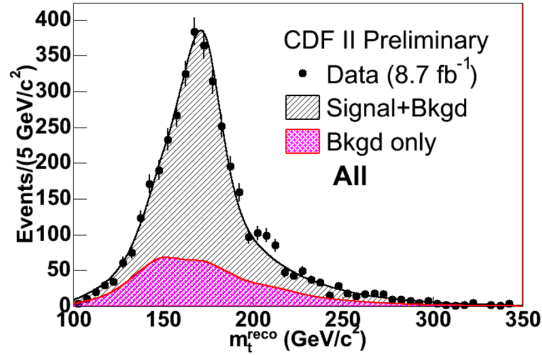}}
\caption[*]{ CDF top mass fit in the lepton+jets channel using the
template method.  The figure is from Reference~\cite{CDF10761}.}
\label{fig:MtCDF}
\end{figure}

The CDF collaboration has recently released a
measurement in 8.7 fb$^{-1}$ of $p\bar{p}$ collisions at $\sqrt{s}$
in the lepton+jets channel using a template fitting
technique~\cite{CDF10761}.  A $\chi^2$ minimization is used to
assign measured objects to final-state partons.  The reconstructed
top mass is a free parameter in the $\chi^2$, with the constraint
that $m_t^{reco}=m_{\bar{t}}^{reco}$, and the $\chi^2$ includes
resolution terms that allow the measured $p_T$ for each object to
vary.  The $\chi^2$ is minimized with respect to $m_t$ and the
distribution of $m_t^{reco}$ values over the ensemble of events is
fit to Monte Carlo templates for the distribution of $m_t^{reco}$ as
a function of the true top mass. The JES uncertainty is controlled
by a simultaneous fit of $m_{jj}$ from the hadronically decaying $W$
boson to the known $W$ boson mass. The result, shown in
Figure~\ref{fig:MtCDF}, is
\begin{equation*}
      m_t=172.9\pm 0.7 {\rm ~(JES+stat.)} \pm 0.8 {\rm ~(syst.)
      ~GeV}
\end{equation*}

The uncertainty due to JES is included as a statistical uncertainty
here due to the $m_{jj}$ fit.  This is the most precise single
measurement of the top-quark mass to date.

\subsection{D0 \ Measurement in Dileptons}
\label{sec:D0DIL} The D0 collaboration has recently measured the top
mass in 5.3 fb$^{-1}$ in the dilepton channel ($ee$, $\mu\mu$,
$e\mu$)~\cite{D0MtDIL}.  In this channel there are two unobserved
neutrinos, resulting in a kinematically under-constrained system. To
address this problem three additional constraints are needed. Two
are given as follows:
\begin{eqnarray*}
 m(\ell_1\nu_1) &=& m(\ell_2\nu_2)=m_W \\
 m(\ell_1\nu_1b_1) &=& m(\ell_2\nu_2b_2)
\end{eqnarray*}

\begin{figure}[!thb]
 \centerline{\epsfxsize=2.9in\epsfbox{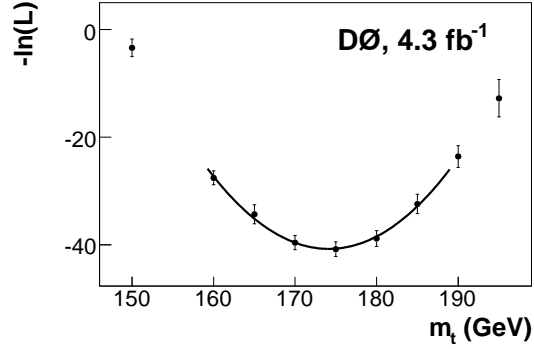}}
\caption[*]{ D0 top mass fit in the dilepton channel using the
neutrino weighting method.  The figure is from
Reference~\cite{D0MtDIL}. The result quoted in the text is a
combination with an earlier 1 fb$^{-1}$ analysis.}
\label{fig:D0MtDIL}
\end{figure}

There are a number of different choices for imposing the third
constraint. In this measurement an integration is performed over the
pseudorapidity of both neutrinos, $\eta_1$ and $\eta_2$.  At each
$(\eta_1,\eta_2)$ the neutrino momenta are calculated and from them
the missing $E_T$. A weight is assigned based on the agreement
between the calculated and measured missing $E_T$, and the missing
$E_T$ resolution.  Figure~\ref{fig:D0MtDIL} shows the maximum
likelihood fit to the weight distribution from which the top mass is
extracted. The result is
\begin{equation*}
    m_t=174.0\pm 2.4 {\rm~(JES+stat.)} \pm 1.4 {\rm~(syst.) ~GeV}
\end{equation*}

\subsection{Tevatron Top Mass Combination}
\begin{figure}[!thb]
 \centerline{\epsfxsize=4.0in\epsfbox{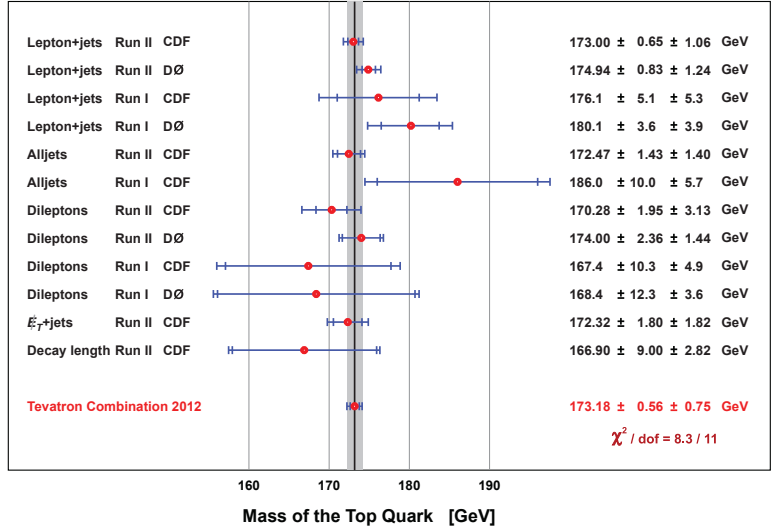}}
\caption[*]{ The twelve input measurements of $m_t$ and the combined
value from the Tevatron experiments.  The figure is from
Reference~\cite{ref:MtMw}.} \label{fig:MtCombo}
\end{figure}

The CDF and D0 collaborations have combined a variety of different
measurements of the top quark mass using integrated luminosities up
to 5.8 fb$^{-1}$~\cite{ref:MtMw}.  The combination does not yet
include the latest results, such as those described in
Sections~\ref{sec:CDFLJ}. Figure~\ref{fig:MtCombo} shows the
measurements used in the combination.  Taking into account
correlations in the systematic uncertainties between the different
measurements, the combined top mass value is found to be
\begin{equation*}
    m_t=173.18\pm 0.56 {\rm~(stat.)} \pm 0.75 {\rm~(syst.) ~GeV}.
\end{equation*}

The top mass measured from Tevatron data has a precision of 0.54\%.
A remarkable achievement.

\subsection{ATLAS \& CMS Top Mass Measurements}
\begin{figure}[!thb]
 \centerline{\epsfxsize=2.7in\epsfbox{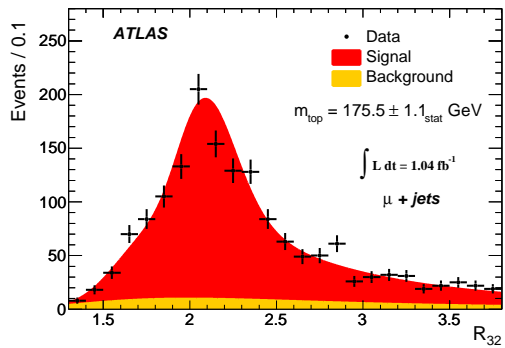}}
\caption[*]{ The $R_{32}$ distribution from ATLAS in the $\mu$+jets
channel. The crosses show the data with its statistical uncertainty,
and the underlying distribution shows the signal and background
contributions determined by the fit.  The figure is from
Reference~\cite{ATLASMt}.} \label{fig:R32}
\end{figure}

At the LHC at $\sqrt{s}=7$ TeV the $t\bar{t}$ production cross
section is roughly 20 times what it is at the Tevatron. Both ATLAS
and CMS have taken advantage of the large number of top quarks to
employ new techniques for measuring $m_t$.
 Using
lepton+jets events, the ATLAS collaboration uses the three jets from
hadronic-side top decay to form the per-event ratio
$R_{32}$~\cite{ATLASMt}, defined as

\begin{equation*}
    R_{32}=\frac{m(qqb)}{m(qq)}
\end{equation*}
The jet triplet ($qqb$) assigned to the hadronic top decay is
selected by maximizing a kinematic likelihood.  The two jets
assigned to the $W$ decay ($qq$) are constrained to $m_W$ in the
likelihood and the JES uncertainty is controlled through the ratio.
The top mass is extracted through a maximum likelihood fit of
$R_{32}$ templates determined from Monte Carlo simulation for a
range of $m_t$ values. Figure~\ref{fig:R32} shows the template fit
result in the $\mu$+jets channel in 1.04 fb$^{-1}$ of data at
$\sqrt{s}$=7 TeV.

\begin{figure}[!thb]
 \centering \subfigure[]{
\includegraphics[width=0.35\textwidth]{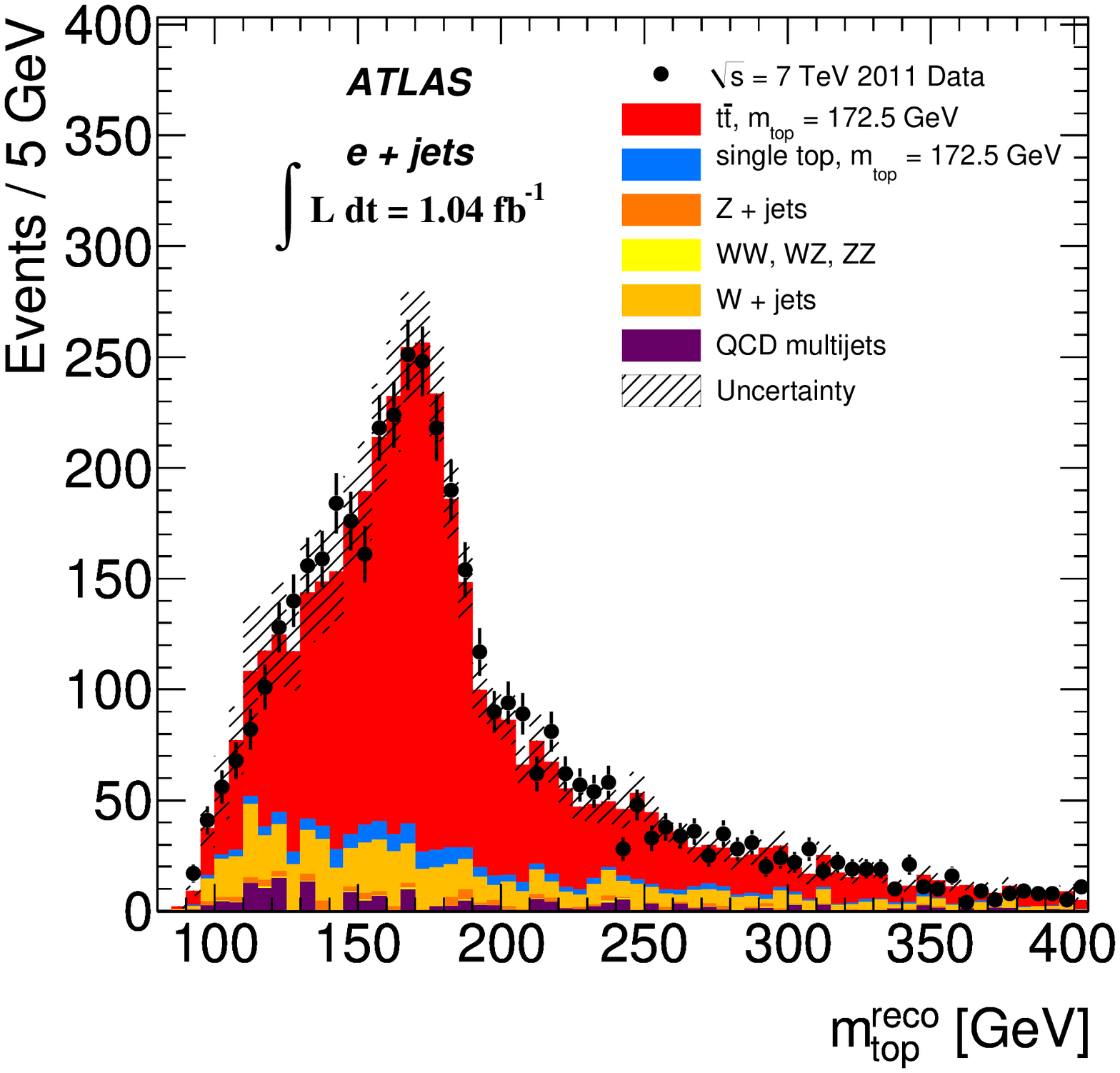}
} \centering \subfigure[]{
\includegraphics[width=0.45\textwidth]{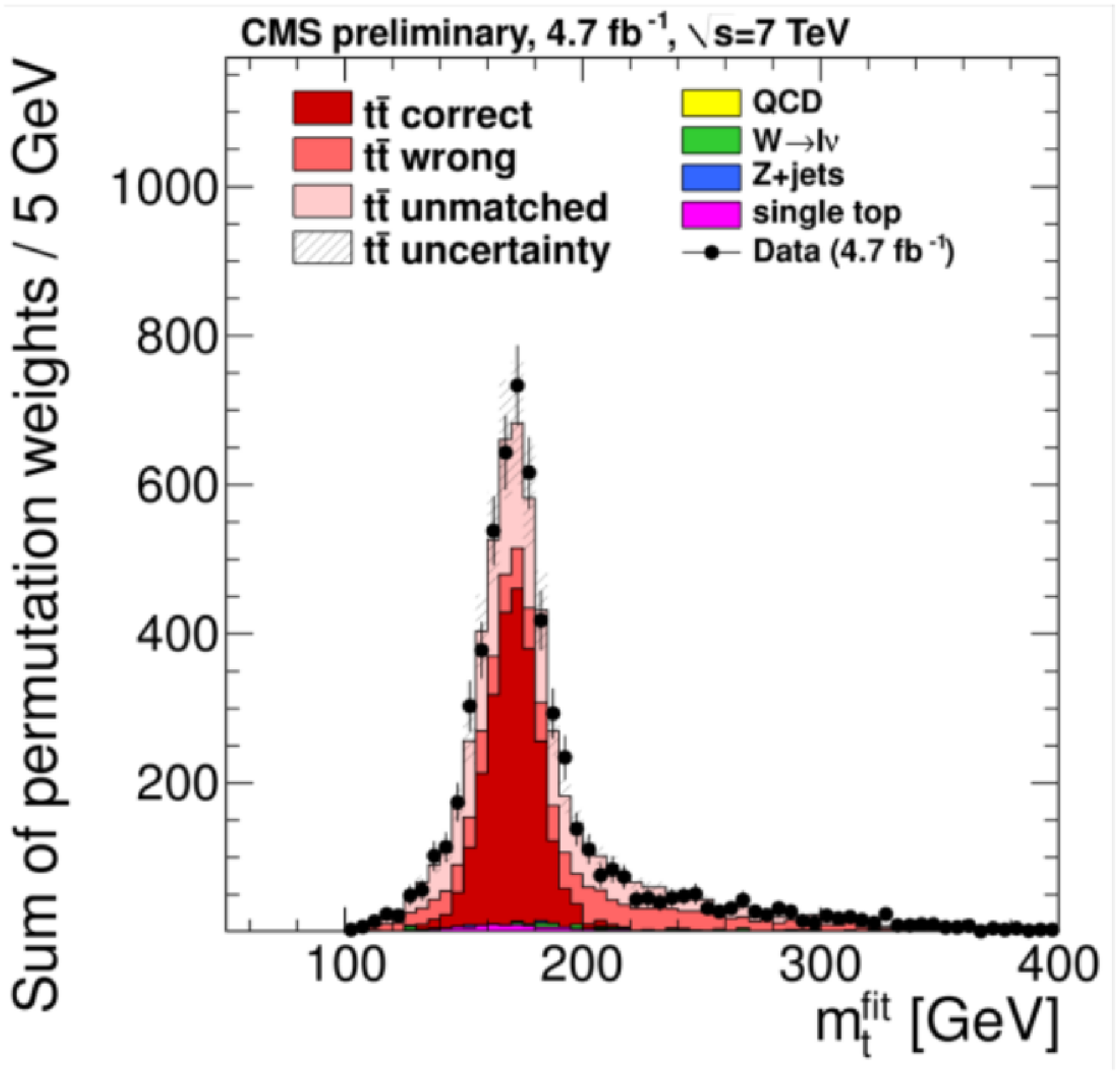}
}
 \caption[*]{ Reconstructed $m_t$ distribution for the {\em best}
$qqb$ assignment from ATLAS (L), and for {\em all} $qqb$ assignments
from CMS (R). The figures are from References~\cite{ATLASMt}
and~\cite{CMSMt}.} \label{fig:CMSATLASMt}
\end{figure}

An alternate ``2-D" method is used by both ATLAS~\cite{ATLASMt} and
CMS~\cite{CMSMt}. In the 2-D method $m(qqb)$ and $m(qq)$ are
measured simultaneously (instead of only in ratio).  ATLAS does a
simultaneous template fit to the {\em best} assignment, defined as
the highest $p_T$ triplet made from a $b$-tagged jet and light-jet
pair with invariant mass between 50 GeV and 100 GeV. In contrast,
CMS uses an ideogram method in which {\em all} jet-parton pairings
are used with a weight given by the probability, from Monte Carlo
simulation, of the pairing being correct.  In both cases the JES is
determined by fitting $m(qq)$ to $m_W$.  The distributions of
reconstructed $m_t$ are fit to templates.  The results are shown in
Figure~\ref{fig:CMSATLASMt}.

ATLAS, in 1.04 fb$^{-1}$ of data at $\sqrt{s}=7$ TeV measures\\
\begin{equation*}
m_t=174.5\pm 0.6 {\rm ~(stat.)} \pm 2.3 {\rm ~(syst.)~GeV},
\end{equation*}
 and CMS in 4.7 fb$^{-1}$ of data at $\sqrt{s}=7$~TeV measures
\begin{equation*}
m_t=172.6\pm 0.6 {\rm ~(stat.)}\pm 1.2 {\rm ~(syst.)~GeV}.
\end{equation*}

\begin{figure}[!thb]
 \centerline{\epsfxsize=4.0in\epsfbox{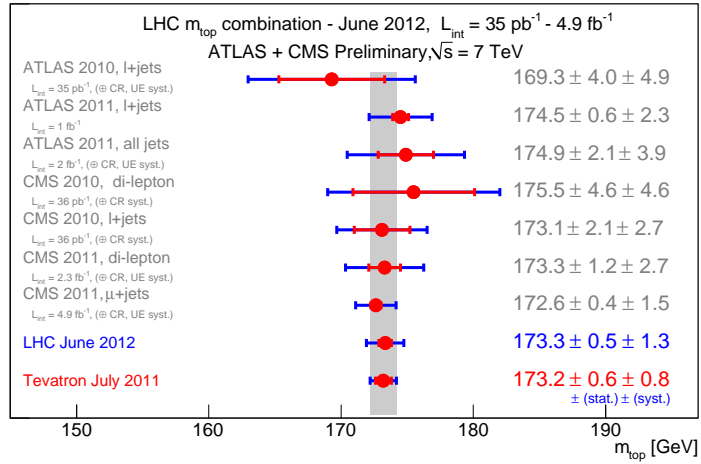}}
\caption[*]{ Top mass measurements from the LHC.  These results use
up to 4.9 fb$^{-1}$ of data taken at $\sqrt{s}=7$ TeV. The figure is
from Reference~\cite{LHCMt}, which does not contain the latest
results from References~\cite{AllJetsMt} and~\cite{DILMt}.}
\label{fig:LHCMt}
\end{figure}

There are additional measurements from ATLAS and CMS, including a
measurement in the all-jets channel from ATLAS~\cite{AllJetsMt}, and
one in the dilepton channel from CMS~\cite{DILMt}. A compilation of
the LHC measurements of $m_t$ is shown in Figure~\ref{fig:LHCMt}.
The combined $m_t$ value from the LHC is
\begin{equation*}
m_t=173.2\pm 0.5 {\rm ~(stat.)}\pm 1.3 {\rm ~(syst.)~GeV}.
\end{equation*}

Separating the lepton+jets data into $\mu^+$ and $\mu^-$ events, CMS
has measured the top-antitop mass difference~\cite{CMSDMt}, which
should be zero if CPT is a good symmetry.  The result is
\begin{equation*}
\Delta m_t = -0.44\pm 0.46 {\rm ~(stat.)} \pm 0.27 {\rm ~(syst.)
~GeV}
\end{equation*}

This is the most precise measurement of the mass difference to date.
It was first measured by D0~\cite{D0DMt} and CDF~\cite{CDFDMt}.

\section{Top Quark Width and $V_{tb}$}
The production cross section for single-top events is sensitive to
the partial width of the top-quark width, $\Gamma(t\rightarrow Wb)$.
The D0 collaboration~\cite{D0tWidth} has combined the measured
$t-$channel single-top production cross section and the ratio of
top-quark branching fraction, $R=B(t\rightarrow Wb)/B(t\rightarrow
Wq)$ to extract the total width of the top quark. The result
\begin{equation*}
\Gamma_t = 2.00^{+0.47}_{-0.43} {\rm GeV},
\end{equation*}
for a top mass of 172.5~GeV, is in good agreement with the Standard
Model value of 1.3~GeV. The measured result is translated into a
top-quark lifetime of $\tau_t =(3.29^{+0.90}_{-0.63})\times
10^{-25}$s.

Since the CKM element $V_{tb}$ and the partial width are
proportional to one another, $V_{tb}$ can be extracted by comparison
of the measured and predicted (using $V_{tb}\simeq 1.0$) single-top
production cross sections.  This has been done by all four
experiments~\cite{Vtb} with results consistent with 1.0 and
precision of order 10\%.

\section{Top-Quark Couplings}
In this section I discuss recent results on the top quark electric
charge, couplings to $W$ and $Z$ bosons, the $W$ helicity in
top-quark decays and searches for flavor-changing-neutral-currents.

\subsection{Top-Quark Electroweak Couplings}
The top-quark electroweak couplings include those to photons and $W$
and $Z$ bosons.  The coupling to photons is sensitive to the
electric charge of the top-quark (which is also measured directly at
both the Tevatron and LHC using jet-charge techniques).  In 1.04
fb$^{-1}$ of data at $\sqrt{s}=7$ TeV, ATLAS has measured the cross
section for $t\bar{t}\gamma$ production~\cite{ttgamma} with a photon
$p_T$ threshold of 8 GeV.  The result is
$\sigma_{t\bar{t}\gamma}=2.0\pm 0.5~{\rm stat.}\pm 0.7~{\rm
syst.}\pm 0.8~{\rm lumi.}$ pb. This compares well with the Standard
Model expectation of $2.1\pm 0.4$ pb for a charge 2/3 top-quark.
There are also direct measurements of the top-quark charge using
jet-charge techniques.  These are discussed elsewhere in these
proceedings~\cite{Federic}.

\begin{figure}[!thb]
 \centerline{\epsfxsize=2.0in\epsfbox{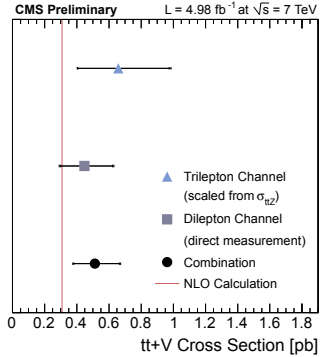}}
\caption[*]{ Measurement of the $t\bar{t}V$ production cross
section: the cross section from the trilepton channel (blue
triangle), from the same sign dilepton channel (grey square) and the
combination of the two measurements (black circle) are compared to
the NLO calculation (red line). Internal error bars for the
measurements represent the statistical component of the uncertainty.
The figure is from Reference~\cite{ttV}.} \label{fig:ttV}
\end{figure}

The associated production of $t\bar{t}$ and $W$ or $Z$ bosons has
been measured by CMS using trilepton events ($t\bar{t}Z$) and
same-sign dilepton events ($t\bar{t}Z$, $t\bar{t}W$) in 4.98
fb$^{-1}$ of pp collisions at $\sqrt{s}=7$TeV~\cite{ttV}. The
results, compared with the Standard Model predictions, are shown in
Figure~\ref{fig:ttV}.

\subsection{$W$-boson Helicity}
Measurements of the $W$-boson helicity in top quark decays probe the
structure of the $Wtb$ vertex which, in the Standard Model, is
$V-A$.  For the Standard Model coupling, the fraction of
longitudinally polarized $W$ bosons is a function of the top quark
and $W$ boson masses, given in the limit of $m_b=0$ by
\begin{equation*}
F_0 =\frac{m_t^2/2m_W^2}{1+m_t^2/2m_W^2}
\end{equation*}
which gives 69.9\% for $m_t=173.3$ and $m_W=80.399$.  The remainder
is almost entirely left-handed.  Even with a non-zero $m_b$ the
right-handed fraction is tiny.

\begin{figure}[!thb]
 \centerline{\epsfxsize=2.5in\epsfbox{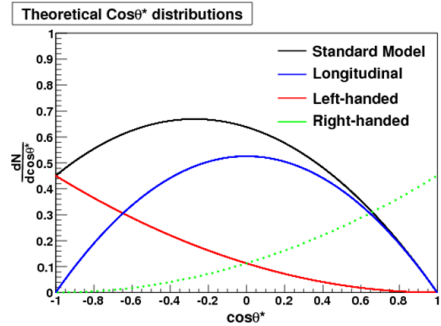}}
\caption[*]{Angular distributions from longitudinal, right-handed
and left-handed $W$ bosons.  The figure is courtesy of CDF.}
 \label{fig:costh}
\end{figure}

The measurement of the helicity fractions is based on the angular
distribution of the $W$ boson decay products. Figure~\ref{fig:costh}
shows the ideal angular distributions for the three helicity states
and their expected Standard Model sum. The angle $\cos\theta^*$ is
the charged lepton angle with respect to the top-quark direction in
the $W$ rest frame.

The measurement can be made in all top-quark decay channels and is
done using a variety of techniques.  D0 uses a template fit to the
reconstructed angular distributions in a combination of lepton+jets
and dileptons $t\bar{t}$ decays in 5.4 fb$^{-1}$~\cite{D0Whel}.  CDF
has used a matrix element technique in 8.7 fb$^{-1}$~\cite{CDFWhel}
of lepton+jets events, that uses the full event kinematics to
calculate an event probability as a function of the longitudinal and
right-handed helicity fractions.  At the LHC, both CMS and ATLAS
have made measurements using template fits to the $\cos\theta^*$
distributions in 2.2 fb$^{-1}$~\cite{CMSWhel} of lepton+jets events
and 1.04 fb$^{-1}$~\cite{ATLASWhel} of lepton+jets and dilepton
events, respectively.  $W$ helicity measurements are made either as
two-parameter fits, or one-parameter fits with one helicity fraction
fixed to its Standard Model value and $F_0+F_R+F_L=1.0$ assumed. A
selection of recent results are
summarized in Table~\ref{tab:Whel}.\\
\begin{table}
\begin{center}
\begin{tabular}{|l|l|l|c|c|}
                       \hline
                       Experiment & Helicity    & Result    &  Channel & Reference\\ \hline
                       D0 & $F_0$ & $0.669\pm 0.078\pm 0.065$ & Dil \& LJ &\cite{D0Whel} \\
                        & $F_R$ & $0.023\pm 0.041\pm 0.034$ & &  \\ \hline
                       CDF & $F_0$ & $0.726\pm 0.066\pm 0.067$ & LJ &\cite{CDFWhel} \\
                        & $F_R$ & $0.045\pm 0.043\pm 0.058$ & & \\ \hline
                       CMS & $F_0$ & $0.567\pm 0.074\pm 0.047$  & LJ &\cite{CMSWhel} \\
                        & $F_L$ & $0.393\pm 0.045\pm 0.029$ & & \\ \hline
                       ATLAS & $F_0$ & $0.67\pm 0.03\pm 0.06$ & Dil \& LJ &\cite{ATLASWhel} \\
                        & $F_L$ & $0.32\pm 0.02\pm 0.03$ & & \\
                        & $F_R$ & $0.01\pm 0.01\pm 0.04$ & & \\ \hline
                       \hline
                       \end{tabular}
                       \caption{Recent measurements of the $W$ boson helicity fractions in top-quark
                       decays.  The notation for the $t\bar{t}$ decay channels is `Dil' for the dilepton
                       channel and `LJ' for the lepton+jets channel. The uncertainties listed are statistical
                       and systematic, respectively.}
                       \label{tab:Whel}
                       \end{center}
                       \end{table}

\subsection{Searches For Flavor-Changing Neutral Currents}
\begin{figure}[!thb]
\centerline{\epsfxsize=3.0in\epsfbox{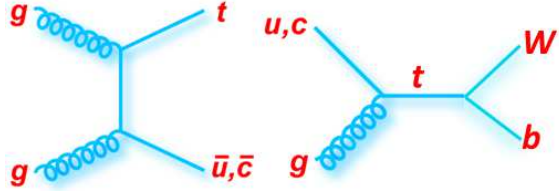}} \caption[*]{Feynman
diagrams for single-top quark production via flavor-chaning neutral
currents.}
 \label{fig:FCNCProd}
\end{figure}

In the Standard Model Flavor-changing neutral current (FCNC)
interactions are forbidden at tree level and highly supressed at
higher order by the GIM mechanism.  Any observation, therefore,
would be a sure sign of physics beyond the Standard Model.  Several
models of physics beyond the Standard Model predict FCNC effects in
top-quark interactions at the $10^{-4}$ level.  The most sensitive
tests in the top-quark sector come from single top-quark production
in which FCNC production diagrams, such as those shown in
Figure~\ref{fig:FCNCProd}, modify the production kinematics.
Searches in single top-quark production are only sensitive to a
$tqg$ coupling.

The first branching fraction limits from single top-quark production
came from CDF~\cite{CDFFCNC} and D0~\cite{D0FCNC}. The most precise
limits which, taking advantage of the relatively large production
cross section at the LHC, come from ATLAS~\cite{ATLASFCNC}, are
\begin{eqnarray*}
  B(t\rightarrow ug)  & < & 5.7\times 10^{-5} \\
  B(t\rightarrow cg) & < & 2.7\times 10^{-4}
\end{eqnarray*}

Direct searches for $t\rightarrow Zq$ FCNC top-quark decays have
been mounted by all four experiments, using final-state kinematics
to search for a signal.  The most precise limits again come from the
LHC, due to the large $t\bar{t}$ production cross section.  Both
CMS~\cite{CMSZq} and ATLAS~\cite{ATLASZq} have searched for
trileptons in $t\bar{t}$ lepton+jets candidate events, with a
same-flavor, opposite-sign pair consistent with a
$Z\rightarrow\ell^+\ell^-$ decay.  In 2.1 fb$^{-1}$ ATLAS sets a
limit of $B(t\rightarrow Zq)<0.73\%$.  In 4.6 fb$^{-1}$, the CMS
limit is $B(t\rightarrow Zq)<0.34\%$.

\section{Top Quark Production Properties}
The production cross section for $t\bar{t}$ pairs and for single
top-quark production are covered elsewhere in these
proceedings~\cite{XsecTalk}, as is the $t\bar{t}$ charge
asymmetry~\cite{AsymTalk}.  Here I concentrate on the top quark spin
direction at production.

The top-quark lifetime is so short that, unlike other quarks, it
decays before hadronization.  As a result, its spin direction at
production is preserved through its decay and measurable through the
angular distributions of its decay products.

\subsection{Top Quark Polarization}
\begin{figure}[!thb]
\centerline{\epsfxsize=3.0in\epsfbox{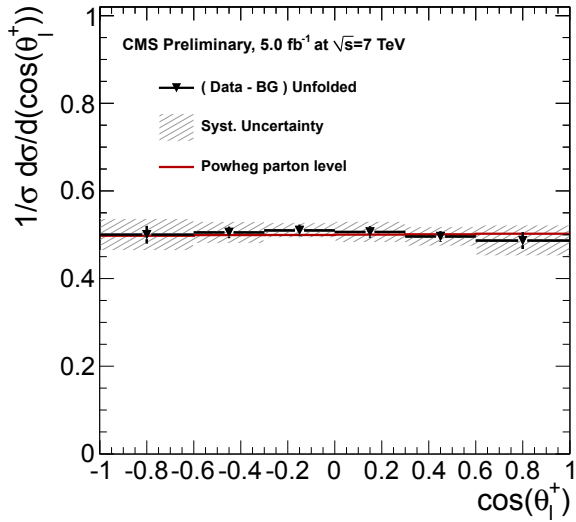}}
\caption[*]{Background-subtracted and unfolded
$\cos(\theta_{\ell}^+)$ distribution with statistical error bars
only.  The systematic uncertainty is given by the hatched area. The
bin values are correlated due to the unfolding. The figure is from
Reference~\cite{CMSPol}.}
 \label{fig:TopPol}
\end{figure}

The production of $t\bar{t}$ pairs occurs through the strong
interaction and therefore, apart from small electroweak corrections,
the top-quark spin is expected to be unpolarized. The first study of
top-quark polarization was made by D0 at the Tevatron~\cite{D0Pol}.
The CMS experiment has recently measured the polarization of
top-quark pairs produced at the LHC~\cite{CMSPol} using 5.0
fb$^{-1}$ of dilepton events. The polarization angle,
$\theta^+_{\ell}$, is defined as the angle between the direction of
the positively-charged lepton, in the rest frame of its parent top
quark, and the parent top-quark direction in the $t\bar{t}$ rest
frame.  The measurement of the polarization angle requires a
reconstruction of the $t\bar{t}$ system, which is done using an
analytic technique.  The background subtracted, unfolded
distribution is shown in Figure~\ref{fig:TopPol}.

The polarization, $P_n$ is calculated from the asymmetry of the
number of events with positive vs. negative values of
$\cos\theta_{\ell}^+$ and found to be
\begin{equation*}
P_n = -0.009\pm 0.029\pm 0.041.
\end{equation*}

\subsection{$t\bar{t}$ Spin Correlations}
While the top-quark spins are unpolarized, they are expected to be
correlated.  At the Tevatron, where the dominant production mode for
$t\bar{t}$ pairs is $q\bar{q}$ annihilation, $t$ and $\bar{t}$ are
expected to be produced mostly with opposite helicities.  At the
LHC, where the production mode is dominantly gluon-gluon
annihilation, the $t\bar{t}$ pairs are expected to be like-helicity
at low $m_{t\bar{t}}$ and opposite-helicity at high $m_{t\bar{t}}$.

The first results have come from the Tevatron.  CDF measures the
correlation in the beam basis, where the angles $\theta^+$ and
$\theta^-$ are defined as the positive and negative lepton
directions with respect to the beam direction in the parent top rest
frame.  Here the double-differential angular distribution is given
by
\begin{equation*}
\frac{1}{\sigma}
\frac{d^2\sigma}{d\cos\theta^+d\cos\theta^-}=\frac{1+\kappa\cos\theta^+\cos\theta^-}{4}
\end{equation*}
CDF uses 5.1 fb$^{-1}$ of dilepton events~\cite{CDFSpin} and a
maximum likelihood technique to reconstruct the events, to measure
$\kappa = 0.043^{+0.563}_{-0.562}$.

D0 uses a matrix element technique in 5.3 fb$^{-1}$~\cite{D0Spin} in
which a discriminant $R$ is calculated for each event, where $R=\wp
({\rm corr})/[\wp ({\rm corr})+\wp ({\rm uncorr})]$ and $\wp ({\rm
corr,uncorr})$ are the matrix-element probabilities for correlated
and uncorrelated spins. The distribution of $R$ is fit with
templates for correlated and uncorrelated spins to find a fraction
of correlated spins of $f=0.85\pm 0.29$, yielding the first
3.1$\sigma$ evidence for spin correlations.  Earlier studies from D0 can be found in Ref.~\cite{D0OldSpinCorr}.

\begin{figure}[!thb]
\centering \subfigure[]{
\includegraphics[width=0.33\textwidth]{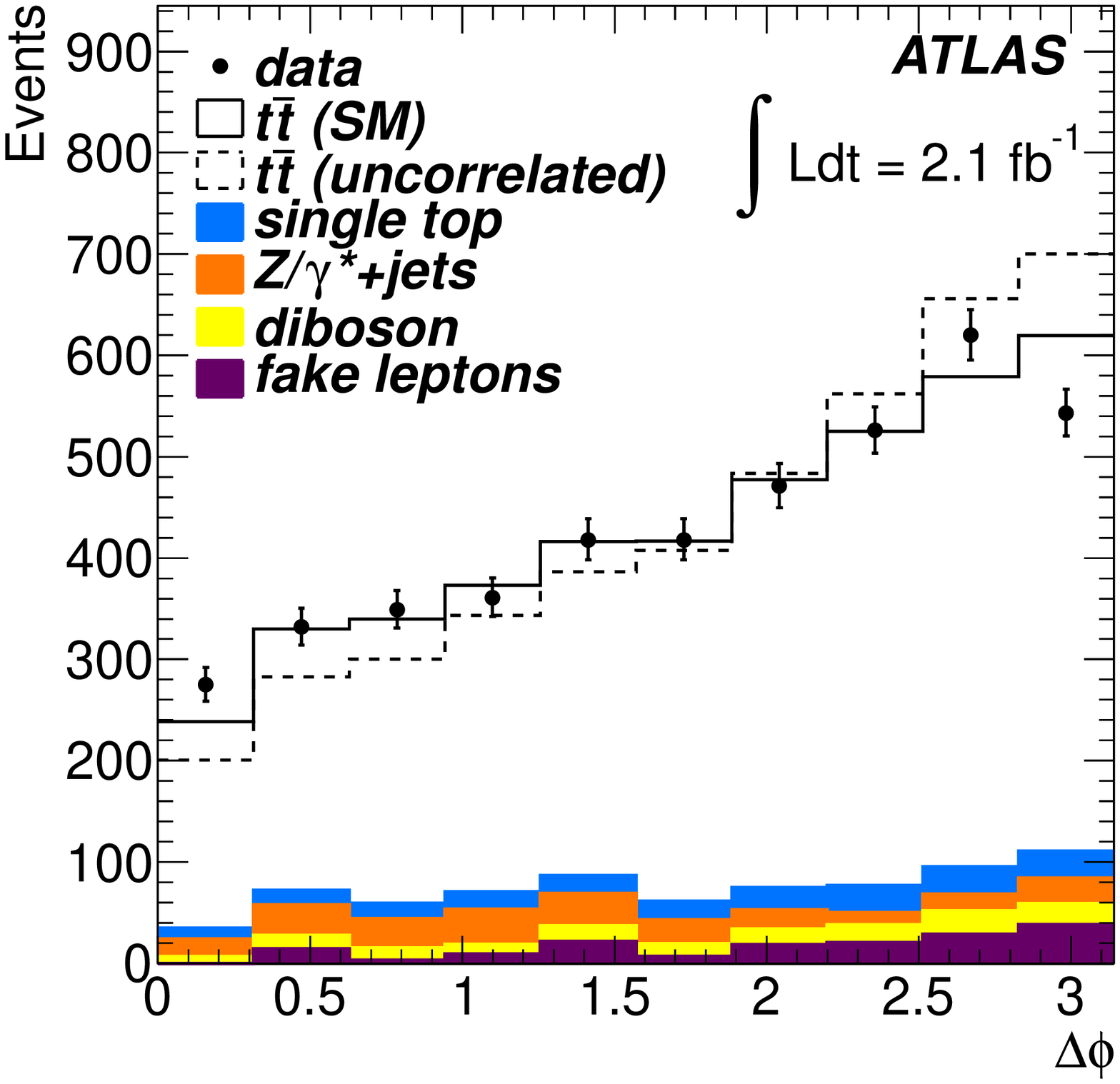}
} \centering \subfigure[]{
\includegraphics[width=0.42\textwidth]{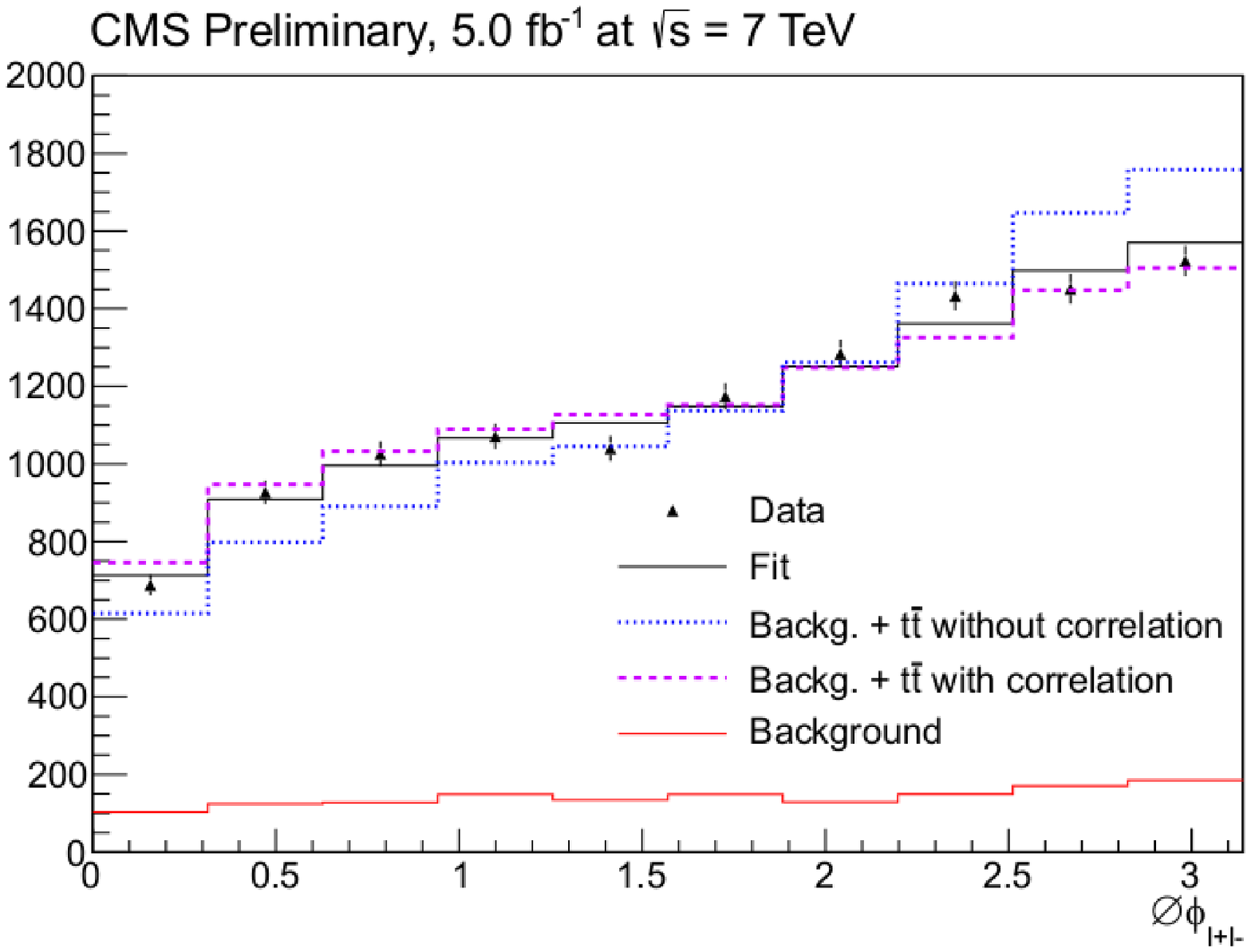}
}
 \caption[*]{
Azimuthal separation between the two leptons in $t\bar{t}$ dilepton
decays from ATLAS (a) and CMS (b). The figures are from
References~\cite{ATLASDPhi} and~\cite{CMSDPhi}.}
\label{fig:DeltaPhi}
\end{figure}

At the LHC, the dominance of the gluon-gluon fusion production of
$t\bar{t}$ pairs provides a straightforward technique for measuring
the spin correlation. In production via gluon-gluon fusion, the spin
correlation is encoded in the angular separation of the two leptons,
$\Delta\phi_{\ell^+ \ell^-}$, in dilepton decays of the $t\bar{t}$
pair~\cite{MahlonParke}.  Figure~\ref{fig:DeltaPhi} shows the data
and fits from both ATLAS and CMS.

The data are fit to a sum of the expected distribution for
correlated spins and the expected distribution for uncorrelated
spins.  The ATLAS result is $f=1.30\pm 0.14^{+0.27}_{-0.22}$, where
the first uncertainty is statistical and the second systematic, and
$f$ is the fraction of events with a Standard Model spin
correlation.  The result excludes $f=0$ at the $5.1\sigma$ level.
CMS measures $f=0.74\pm 0.08\pm 0.24$. CMS has also searched for
signs of new phyiscs in the dependence of $\Delta\phi_{\ell^+
\ell^-}$ on $m_{t\bar{t}}$.  Data at both low and high ($>450$ GeV)
$m_{t\bar{t}}$ are consistent with Standard Model expectations.  CMS
also presents results on the asymmetry $A_{c1c2}$ defined as the
fraction of events with the product $\cos\theta^+\times\cos\theta^-$
positive minus the fraction with that product negative.  This result
requires reconstruction of the $t\bar{t}$ frame in dilepton events
and unfolding to correct for detector effects.  It is, however, also
sensitive to spin correlation in $q\bar{q}$ production of $t\bar{t}$
pairs.  While less precise than the measurement of $f$, the measured
value is also consistent with Standard Model expectations.

\section{Conclusions}
With as much as 9 fb$^{-1}$ of data from the Tevatron, and as much
as 5 fb$^{-1}$ analyzed and at least three times that much still to
come from the LHC, the era of precision top physics is upon us.  I
have reviewed the latest results from the four experiments on the
top-quark mass, couplings, and searches for rare decay modes.  All
results are, so far, consistent with Standard Model predictions.  As
more data are analyzed, and new techniques employed, the precision
of these measurements, and sensitivity to beyond-the-Standard-Model
effects, will continue to increase.

\section*{Acknowledgements} I thank the members of the CDF, D0,
ATLAS, and CMS collaborations for the hard work in producing these
results and the Fermilab and CERN staff whose outstanding work on
the Tevatron and the LHC have produced such rich datasets.  This
work was supported in part by the U.S. Department of Energy.


\end{document}